\documentclass{iau} 
\usepackage{graphicx}

\title[Long-term observations of pulsars in 47 Tuc and M15] 
{Long-term observations of pulsars in the globular clusters 47 Tucanae and M15}

\author[A. Ridolfi et al.]   
{A.\,Ridolfi$^1$, P.\,C.\,C.\,Freire$^1$, M.\,Kramer$^{1,2}$,  C.\,G.\,Bassa$^3$, F.\,Camilo$^4$, N.\,D'Amico$^{5,6}$, G.\,Desvignes$^1$, C.\,O.\,Heinke$^7$, C.\,Jordan$^2$, D.\,R.\,Lorimer$^8$, A.\,Lyne$^2$, R.\,N.\,Manchester$^9$, Z.\,Pan$^{10}$, J.\,Sarkissian$^{11}$, P.\,Torne$^{1,12}$, M.\,van\,den\,Berg$^{13}$, A.\,Venkataraman$^{14}$ and N.\,Wex$^1$ }
\affiliation{
$^{1}$Max-Planck-Institut f\"ur Radioastronomie, Auf dem H\"ugel 69, D-53121 Bonn, Germany \\ email: {\tt ridolfi@mpifr-bonn.mpg.de} \\[\affilskip]
$^{2}$Jodrell Bank Centre for Astrophysics, School of Physics and Astronomy, \\ The University of Manchester, Manchester M13 9PL, UK\\[\affilskip]
$^{3}$ASTRON, the Netherlands Institute for Radio Astronomy, \\Postbus 2, 7990 AA, Dwingeloo, the Netherlands\\[\affilskip]
$^{4}$Square Kilometre Array South Africa, Pinelands, 7405, South Africa\\[\affilskip]
$^{5}$Osservatorio Astronomico di Cagliari, INAF, 	via della Scienza 5, I-09047 Selargius (CA), Italy\\[\affilskip]
$^{6}$Dipartimento di Fisica, Universit\`a degli Studi di Cagliari, \\SP Monserrato-Sestu km 0,7, 90042 Monserrato (CA), Italy\\[\affilskip]
$^{7}$Department of Physics, University of Alberta, CCIS 4-183, Edmonton, AB T6G 2E1, Canada\\[\affilskip]
$^{8}$Department of Physics and Astronomy, West Virginia University, \\P.O. Box 6315, Morgantown, WV 26506, USA\\[\affilskip]
$^{9}$CSIRO Astronomy and Space Science, Australia Telescope National Facility, \\Box 76, Epping, NSW 1710, Australia\\[\affilskip]
$^{10}$National Astronomical Observatories, Chinese Academy of Sciences,\\A20 Datun Road, Chaoyang District, Beijing 100012, China\\[\affilskip]
$^{11}$CSIRO Astronomy and Space Science, Parkes Observatory, \\PO Box 276, Parkes NSW 2870, Australia\\[\affilskip]
$^{12}$Instituto de Radioastronom\'ia Milim\'etrica, \\Avda. Divina Pastora 7, N\'ucleo Central, E-18012 Granada, Spain\\[\affilskip]
$^{13}$Harvard$-$Smithsonian Center for Astrophysics, \\60 Garden Street, Cambridge, MA 02138, USA\\[\affilskip]
$^{14}$Arecibo Observatory, HC3 Box 53995, Arecibo, PR 00612, USA\\[\affilskip]
}

\pubyear{2017}
\volume{337}  
\jname{Pulsar Astrophysics -­ The Next 50 Years}
\editors{P. Weltevrede, B.B.P. Perera, L. Levin Preston \& S. Sanidas, eds.}

\begin{document}

\maketitle
\begin{abstract}
Multi-decade observing campaigns of the globular clusters 47 Tucanae and M15 have led to an outstanding number of discoveries.
Here, we report on the latest results of the long-term observations of the pulsars in these two clusters. For most of the pulsars in 47 Tucanae we have measured, among other things, their higher-order spin period derivatives, which have in turn provided stringent constraints on the physical parameters of the cluster, such as its distance and gravitational potential.  For M15, we have studied the relativistic spin precession effect in PSR B2127+11C. We have used full-Stokes observations to model the precession effect, and to constrain the system geometry. We find that the visible beam of the pulsar is swiftly moving away from our line of sight and may very soon become undetectable. On the other hand, we expect to see the opposite emission beam sometime between 2041 and 2053.

\end{abstract}

\firstsection 
\section{Introduction}
Globular clusters (GCs) are spherical, self-gravitating aggregations of stars orbiting the bulge of a host galaxy. In the Milky Way, 157 GCs are currently known (\cite{Harris1996}, version 2010). With typical ages of $\sim$10 billion years, GCs harbor a large number of neutron stars, some of which are observable as radio pulsars.  
30 years since the discovery of the first GC pulsar (PSR B1821$-$24A; \cite{Lyne+1987}), we have now discovered 149 pulsars in 28 different clusters\footnote{http://www.naic.edu/$\sim$pfreire/GCpsr.html}. Such a population, however, is not equally distributed: more than 50\% of the known pulsars are hosted by just four clusters. Among these are 47 Tucanae and M15. Hosting 25 and 8 known pulsars respectively, these two clusters have proved to be remarkably fertile grounds for a number of scientific feats.
In the following, we report on the newest results obtained from multi-decade observations of both clusters.

\vspace{-0.3 cm}

\section{47 Tucanae}
47 Tucanae (henceforth, 47 Tuc) is a GC located in the Southern sky. Since 1989, it has been regularly observed with the Australian 64-m Parkes radio telescope. Our data consisted of 519 pointings of the cluster made from October 1997 to August 2013. The majority of the observations were made with the Parkes Multibeam Receiver (PMB) at a central frequency of $\sim$1.4 GHz, with two different backend systems: the PMB filterbank, which allowed 288 MHz of bandwidth, divided into 96, 3-MHz wide channels;  and the Analogue Filterbank (AFB), with 256 MHz of bandwidth, divided into 512, 0.5-MHz wide channels. The data were first searched with a new incoherent stack-search technique (\cite{Pan+2016}), which resulted in the discovery of two new isolated pulsars, 47 Tuc aa and ab. A standard acceleration search was then used to maximize the detections of all pulsars in the cluster. The new detections allowed us to derive phase-connected timing solutions for the two new pulsars, as well as for another five previously known, but unsolved pulsars (47 Tuc R, W, X, Y, Z; \cite{Ridolfi+2016,Freire+2017}). For the other 18 pulsars, we have extended their timing solutions by adding about 10 more years of data, compared to their last publication by \cite{Freire+2003}.
The total number of pulsars in 47 Tuc with phase-connected timing solutions is now 23. The remaining two pulsars (47 Tuc P and V) are extremely faint, and were detected too few times for a timing solution to be found. Nonetheless, as both are in extremely compact binary systems, we were able to greatly refine the measurements of their orbital parameters. 
Having timing solutions that span more than two decades is important for these systems, as it allows to observe peculiar properties of each pulsar that become measurable only with long baselines. For example, we have measured the rate of advance of periastron, $\dot{\omega}$, in three binary systems: 47~Tuc~E ($\dot{\omega}\, = \, 0.090 \, \pm \, 0.016 ~{\rm deg}~\rm yr^{-1}$), 47~Tuc~S ($\dot{\omega}\, = \, 0.311 \, \pm \, 0.075 ~{\rm deg}~{\rm yr}^{-1}$), and  47~Tuc~U ($\dot{\omega}\, = \, 1.17 \, \pm \, 0.32 ~{\rm deg}~{\rm yr}^{-1}$). Another example is the detection of a strong orbital variability in several black widow and redback pulsars (47 Tuc J, O, V, and W).  Furthermore, we are now able to measure the proper motions of the pulsars relative to each other, to high significance. However, the most interesting scientific results are obtained when using the properties of all pulsars collectively. Thanks to long-term timing, we were able to measure first- and higher-order spin period derivatives for many pulsars in 47 Tuc. The observed spin period derivative $\dot{P}_{\rm obs}$ of a pulsar is the result of several contributions, such as the pulsar proper motion (Shklovskii effect), the acceleration due to the Galactic potential, the acceleration due to the cluster potential and the intrinsic pulsar spin-down ($\dot{P}_{\rm int}$). While the first two contributions can be  directly calculated, the latter two are difficult to independently estimate.
However, if we are able to measure the \emph{orbital} period derivative of a binary system (which is dominated by acceleration in the cluster), then we can directly calculate the acceleration along the line of sight, $a_{\ell, {\rm GC}}$, undergone by the pulsar due to the GC gravitational potential. This is the case for ten binaries in 47 Tuc (pulsars E, H, I, Q, R, S, T, U, X and Y). We compared the observed values of $a_{\ell, {\rm GC}}$ for these pulsars against a simple analytical King model (first used by \cite{Freire+2005}) and found that all of the observed pulsar accelerations can be accounted for by this model, with no need of introducing an intermediate-mass black hole at the center, provided that the cluster distance is $\gtrsim 4.69$~kpc. This is in agreement with published distance estimates for 47 Tuc (e.g. \cite{Woodley+2012}).

\vspace{-0.3 cm}

\section{M15}
The globular cluster M15 has been observed with the 305-m Arecibo radio telescope from April 1989 to August 2017. The 430-MHz line feed was used as the receiving system until 2001, after which virtually all observations were carried out at a central frequency of 1.4 GHz. Most of the data have been taken recording total intensity only. From July 2014, however, all observations have been made using the new Puertorican Ultimate Pulsar Processing Instrument (PUPPI) backend. The latter provided coherently de-dispersed data, retaining full-Stokes information.
\begin{figure}[t]
\begin{center}
 \includegraphics[width=\textwidth]{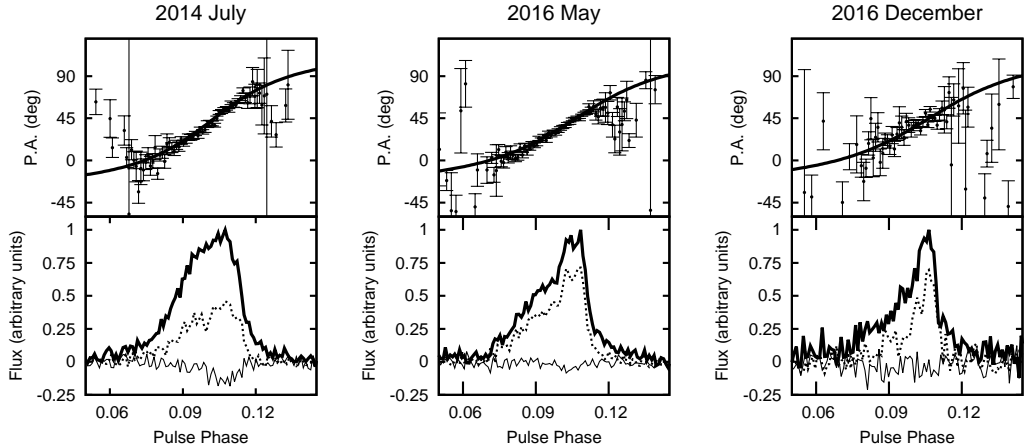} 
 \caption{Evolution of the polarimetric properties of M15C from 2014 to 2016. Bottom: total intensity (thick solid line), linear (dotted line) and circular polarization (thin solid line) profiles. Top: corresponding evolution of the P.A. swing (bars) and  best-fitting solution (thick solid line) of our precessional RVM model.}
   \label{fig1}
\end{center}
\end{figure}
The whole dataset has been used to extend the timing solutions of all pulsars in the cluster. However, in the recent 1.4-GHz data, only five of the eight pulsars were detectable: pulsars A, B, C, D and E.
The most recent full-Stokes PUPPI data were also used to study the polarimetric properties of the latter pulsars.  Polarization calibration, propaedeutic for the correct interpretation of the data, was performed by making a short observation of the receiver's noise diode immediately before or after the observation of the cluster itself. All measurements of the rotation measures (RMs) of the pulsars are consistent within the uncertainties, with an average value of $\sim 70.5$~rad\,m$^{-2}$. Therefore we do not see any significant variation in the RM values, contrary to what has recently been observed in the pulsars in 47 Tuc (see contribution by Abbate et al., this volume). 
The study of the polarimetric properties of the M15 pulsars has largely been motivated by pulsar PSR B2127+11C (henceforth, M15C). This pulsar is part of a relativistic double neutron star system (\cite{Anderson1993}) with a compact (orbital period $P_{\rm b} \sim 8$~hr) and eccentric ($e \sim 0.68$) orbit. The binary had already exhibited three measurable post-Keplerian effects, namely the rate of advance of periastron, the Einstein delay, and the orbital decay due to the emission of gravitational waves (\cite{Jacoby+2006}). More recently, this pulsar has shown a steady decrease in  brightness and significant changes in its pulse profile shape (see Fig. 1). Such variations  (not seen in the other detectable pulsars in our dataset) can be ascribed to relativistic spin precession (RSP, e.g. \cite{Barker_OConnel1975}), an effect predicted by general relativity. Due to RSP, the pulsar spin axis precesses about the total angular momentum vector of the system, with a period of $\sim 275$~yr for  M15C. As a consequence, our line of sight will cut different sections of the emission beam over time, resulting in a slowly time-varying observed profile shape, intensity, and polarimetric properties.
We used our full-Stokes data taken with PUPPI from 2014 to model RSP in M15C. In particular, we fitted the polarimetric data with \texttt{ModelRVM}\footnote{https://github.com/gdesvignes/modelRVM}. The code performs a global fit of the linear polarization position angle (P.A.) values of the considered epochs for the so-called Rotating Vector Model (RVM, \cite{rk1969}) jointly with the model of the pulsar spin precession (\cite{Kramer_Wex2009}), in a self-consistent way. The fit allowed us to constrain the geometry of the system, and we find a pulsar magnetic inclination of $\alpha = 115^{+5}_{-5}$ deg (i.e. we are looking at the pulsar's ``southern beam''),   and a large spin-orbit misalignment angle, $\delta = 76^{+22}_{-20}$~deg. Our line of sight is rapidly moving away from the magnetic axis of the visible beam and, depending on the actual size of the emitting region, the latter might be undetectable in the very near future. Correspondingly, the secondary (or ``northern'') beam, currently not visible, is approaching us and we expect it to become visible sometime between 2041 and 2053. The large uncertainties are mostly due to the very short time span ($\sim 2.5$~yr) of the polarimetric data used, when compared to the precession period. Continued monitoring of the pulsar in the coming months will be essential to improve our model and better constrain the system geometry, before M15C precesses out of sight.

\end{document}